# Spatially controlled fabrication of single NV centers in IIa HPHT diamond

**SERGEI D. TROFIMOV**[1,2,*], **SERGEY A. TARELKIN**[1,2,3], **STEPAN V. BOLSHEDVORSKII**[1,4], **VITALY S. BORMASHOV**[5], **SERGEY YU. TROSHCHIEV**[2,5], **ANTON V. GOLOVANOV**[2], **NIKOLAI V. LUPAREV**[2], **DMITRII D. PRIKHODKO**[1,2], **KIRILL N. BOLDYREV**[6], **SERGEY A. TERENTIEV**[2], **ALEXEY V. AKIMOV**[4], **NIKOLAY I. KARGIN**[7], **NIKOLAY S. KUKIN**[7], **ALEXANDER S. GUSEV**[7], **ANDREY A. SHEMUKHIN**[8], **YURI V. BALAKSHIN**[8], **SERGEI G. BUGA**[2] AND **VLADIMIR D. BLANK**[2]

[1]*Moscow Institute of Physics and Technology, 9 Institutskiy per., Dolgoprudny, Moscow Region 141700, Russia*
[2]*Technological Institute for Superhard and Novel Carbon Materials, 7a Tsentralnaya st., Troitsk, Moscow 108840, Russia*
[3]*National University of Science and Technology MISiS, 4 Leninsky pr., Moscow 119991, Russia*
[4]*Lebedev Physical Institute of the Russian Academy of Sciences, 53 Leninskiy pr., Moscow 119991, Russia*
[5]*The All-Russian Research Institute for Optical and Physical Measurements, 46 Ozernaya st. Moscow 119361, Russia*
[6]*Institute of Spectroscopy of the Russian Academy of Sciences, 5 Fizicheskaya st., Troitsk, Moscow 108840, Russia*
[7]*National Research Nuclear University "Moscow Engineering Physics Institute", 31 Kashirskoye shosse, Moscow 115409, Russia*
[8]*Skobeltsyn Institute of Nuclear Physics (MSU), 1(2) Leninskie gory, Moscow 119991, Russia*
*\*sergey.d.trofimov@phystech.edu*

Single NV centers in HPHT IIa diamond are fabricated by helium implantation through lithographic masks. The concentrations of created NV centers in different growth sectors of HPHT are compared quantitatively. It is shown that the purest {001} growth sector (GS) of HPHT diamond allows to create groups of single NV centers in predetermined locations. The {001} GS HPHT diamond is thus considered a good material for applications that involve single NV centers.

## 1. Introduction

Luminescence centers in diamond are combinations of diamond crystal lattice defects, mainly of impurity atoms and vacancies. Of all color centers, only a few have been observed at a single-defect level. Nitrogen-vacancy (NV) center was the first such object to be explored [1].

The NV center consists of a nitrogen atom in a substitution position and a vacancy in one of four adjacent positions. Optically excited negatively charged $NV^-$ centers (referred to as NV hereinafter) luminesce in the range of ~ 600-800 nm with the zero phonon line (ZPL) at 637 nm and have spin-dependent luminescence and long coherence time [2–4]. Single NV centers are considered as logical elements for a quantum processor (quantum registers) or single-photon sources for quantum cryptography [5,6]. In addition, NV centers can be used as nanoscale magnetic field detectors with high spatial resolution [7].

Over the previous twenty years, NV centers have been intensively studied all over the world, but the problem of technologically simple creation of single color centers with good spatial localization has not been completely solved. In addition, studies of NV centers properties were carried out mainly on type Ib diamonds grown under high-pressure high temperature

(HPHT) [8,9], which have high concentrations of nitrogen ($2 \times 10^{17}$–$10^{19}$ cm$^{-3}$ or ~ 1–50 ppm, for diamond 1 ppm = $1.8 \times 10^{17}$ cm$^{-3}$), and high-purity diamonds grown by chemical vapor deposition (CVD) with low nitrogen concentration ($10^{14}$–$10^{17}$ cm$^{-3}$ or 0.5–500 ppb) [10,11]. It is usually supposed that HPHT bulk diamonds are not suitable for single NV centers formation due to high concentration of residual nitrogen and other impurities. However, the properties of NV centers in high-quality IIa HPHT diamond has not yet been intensively studied.

All widely used methods for creating NV centers (ion implantation of nitrogen into a pure crystal [12], implantation of another ion into a nitrogen containing crystal [13], electron irradiation [14]) include annealing at high temperature ($T > 650$ ° C) in order to induce the movement of vacancies in the crystal lattice (except for doping with nitrogen in the process of diamond CVD synthesis [15]). Creation of the NV center is an energetically advantageous process, so vacancies are combined with nitrogen impurity in a substitution position to form an NV center. If there is a sufficient number of donors in the crystal lattice, the NV centers will be mostly negatively charged (in HPHT IIa diamond, the NV$^-$/NV$^0$ ratio is approximately 7:3 [16]). At the same time, the divacancies formation process, competing with the formation of NV centers, occurs [17].

In state-of-art experiments single NV centers are fabricated with high spatial accuracy by single-ion implantation [18], which requires complex and expensive equipment with the ability to create, accelerate and focus single nitrogen ions.

Present work aims to develop a method of controlled creation of single NV centers in high quality low-nitrogen HPHT IIa diamond in predetermined locations.

## 2. Materials and methods

### 2.1. Experimental sample

Bulk type IIa diamond single crystal was grown at the Technological Institute for Superhard and Novel Carbon Materials (TISNCM, Troitsk, Russia) in high-pressure (5 GPa) and high-temperature (~ 1750 K) conditions using the temperature gradient method (TG-HPHT). High-purity (99.9995%) graphite was used as the carbon source and Fe–Al–C alloy (91:5:4 by wt%) as the solvent. Aluminum acts as nitrogen getter and allows to grow colorless IIa type diamonds. Diamond plate of (111) orientation was laser-cut out of grown crystals and mechanically polished to $Ra < 5$ nm.

The luminescence centers were created by ion implantation through the hardmasks patterned with holes followed by high temperature vacuum annealing. The NV03 sample had residual nitrogen concentration of $10^{16}$–$10^{17}$ cm$^{-3}$ (50–500 ppb) depending on growth sector. The concentration data in {111} growth sector were obtained from ultraviolet absorption (see Appendix A) and was ~ $10^{17}$ cm$^{-3}$ (or 500 ppb) [19]. In {001} growth sector the nitrogen content was undetectable via this method. Considering this, we estimated the nitrogen concentration in {001} growth sector to be < $10^{16}$ cm$^{-3}$ or 50 ppb (at least one order lower than in {111}). It is known that due to high growth temperature in TG-HPHT process vacancy complexes are unstable and there are almost no detectable NV centers in as-grown crystals even in {111} growth sector [20]. Based on this, we decided to use the vacancy-driven scheme (Fig. 1(a)) and implanted $^4$He$^+$ ions to locally create additional vacancies.

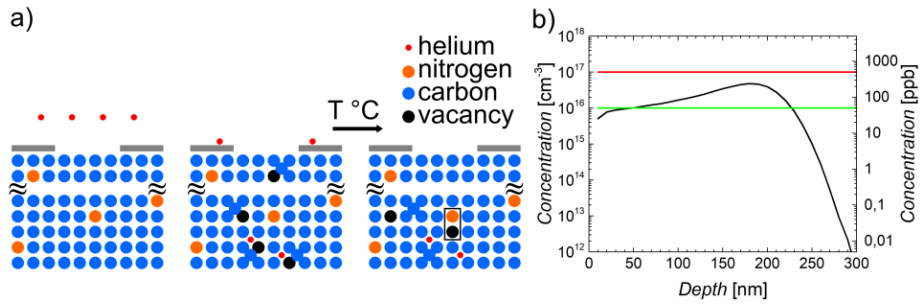

Fig. 1. a) The ion implantation scheme for the NV03 sample (vacancy-driven technique). b) Concentration of vacancies (black) in HPHT diamond after 4He+ implantation. Concentrations of residual nitrogen are shown as horizontal lines (green in {001} and red in {111} sectors).

To determine the optimal energy and dose of implanted ions, we carried out a numerical simulation of the implantation process using Stopping and Range of Ions in Matter (SRIM) software [21]. The proximity of the diamond surface affects the quantum properties of NV centers [22]. We plan to use the created NVs in HPHT diamond as single photon sources located under CVD solid immersion lenses (SIL) or inside a waveguide. A typical height of a waveguide is ~ 500 nm, so it is better to create NVs at a depth of ~ 250 nm [23]. For SIL fabrication the NVs depth must be high enough for NVs not to be etched during CVD SILs overgrowth. The calculated distribution of vacancies is strongly inhomogeneous with a clear maximum. The ion energy determines the depth of the region with the maximum vacancies density and the dose determines the vacancies concentration in this region. According to calculation results, the optimal implantation parameters were: energy – 50 keV and dose – $10^{10}$ ions/cm$^2$ = 100 ions/μm$^2$. Fig. 1(b) shows simulated "frozen" (at T = 0 K) depth profile of vacancies created by He ions implantation. Concentration of residual nitrogen is depicted as a horizontal line.

Before the implantation, we fabricated temperature-resistant coordinate markers in the corners of diamond sample using reactive ion etching in SF6 plasma by method from Ref. [24]. Referencing to the etched markers, we fabricated a metal coordinate grid on diamond surface by lift-off photolithography and magnetron sputtering. The first purpose of the grid was to divide the sample surface into 100 × 100 μm square regions and therefore to simplify the search of the NV centers. The second purpose was to get regularly arranged markers on the surface to focus on them. The width of the grid lines was 10 μm. Optical transparency of diamond makes it very hard to focus on the polished surface when using photolithography. Details of diamond surface preparation and laser photolithography on the small samples can also be found in Ref. [24].

Next, we fabricated the hardmask for the ion implantation. The 400 nm thick nickel mask was patterned with circle holes from 0.5 to 19.5 μm in diameters.

After the implantation, hardmasks and the grid were removed in aqua regia and sample was annealed in vacuum at 700 °C. Annealing caused migration of vacancies over the crystal and their merging with nitrogen impurities leading to the NV centers formation. Then, referencing to the etched coordinate markers, we fabricated new metal coordinate grids on the sample surface. The second grid layout was the same as of the first one.

Any bulk HPHT sample consists of different growth sectors (GS) (Fig. 2(a)). Growth sector is a crystal region corresponding to crystallographic plane perpendicular to the direction along which new atoms attach to the crystal lattice during the HPHT growth process. Different GSs contain nitrogen in different concentrations [25]. In the {001} GS the concentration of nitrogen (usually 10–100 ppb) is lower than in the {111} GS (100–1000 ppb). GSs can be visualized by ultraviolet photoluminescence (UV PL) method (Figs. 2(a) and 2(b)) or by cathodoluminescence (CL) method [26].

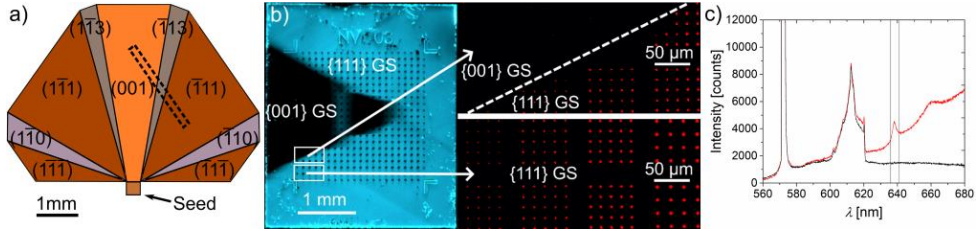

Fig. 2. a) Growth sectors of HPHT diamond. The NV03 plate is depicted by the dashed line. b) NV03 luminescence in UV. {111} GS is blue and {001} GS is black. Luminescence maps of areas in white rectangles are shown in zoom. c) Luminescence spectrum from one of the bright points in Fig. 2(b) (red) and from the dark region (black) in {111} GS. Vertical lines show the integration range for luminescence maps in Fig. 2(b). The peak in range of 600-620 nm corresponds to the diamond second order Raman peak ($\Delta k$ = 2200–2700 cm$^{-1}$ from 532 nm laser line).

## 2.2. Research method

We visualized the location of NV centers using the luminescence mapping on a Renishaw inVia Reflex optical microscope with a 532 nm laser and an air objective (Leica N PLAN EPI 100X NA 0.85). Areas containing NV centers were divided into matrixes of points, in which we measured the luminescence spectrum (Fig. 2(c)). From these data, maps of the integrated luminescence intensity in the range of 636-641 nm (Fig. 2(b)) were derived. The accumulation time was 20 s and averaging over 3 spectra was performed.

To analyze the spatial distribution of the NV centers in irradiated areas we used home-built confocal microscope employing high numerical aperture immersion oil objective (NIKON Apo TIRF 100X NA 1.49) with working distance of 120 micrometers. A continuous wave laser (Coherent Compass 300, wavelength 532 nm) served as the source of NV excitation and galvo mirrors (Cambridge Technologies) were used as a scanning element, enabling up to $100 \times 100$ micrometer field of view for our microscope design (see Fig. 3(a)). The collected fluorescence was coupled into a Hunbury-Brown-Twiss (HBT) interferometer that consisted of two avalanche photodiodes (PerkinElmer SPCM-AQRH-14-FC) and a 45:55 beam splitter (Thorlabs BP145B1). We used the combination of an optical notch filter with a stop-band centered at 532 nm and longpass optical filter with cut-off at 600 nm to remove the residual green excitation light and Raman signal from the collected emission. A time-correlated single photon counting module (Picoquant Picoharp 300) was used to obtain second-order photon correlation histograms from NV centers (see, for example, Fig. 3(b)). The details of background subtraction can be found in Ref. [27]. All measurements were performed at room temperature.

Due to the possible overlapping of NVs in a confocal spot, however, it was impossible to make a confident conclusion about the number of NVs based on a luminescence map without auxiliary experiments.

An important feature of a single luminescence center is the fact that on average some time takes place between emissions, characterizing the lifetime of the system in the excited state (~ 10 ns for the excited state $^3$E of the NV center [6]). For such a system, it is impossible to emit two photons simultaneously. Thus, the second-order (intensity) autocorrelation function $g^{(2)}(t)$ (ACF) of a single NV center should have an antibunching dip to zero at $t = 0$. If the registered photons belong to several luminescence centers located next to each other, the number of these centers can be determined, using the following equation [28]:

$$N = \frac{1}{1 - g^{(2)}(0)},$$

where $N$ is the number of optical active NV centers and $g^{(2)}(0)$ – amplitude of the ACF at $t = 0$.

The concentration of as-grown ("intrinsic") NV centers was measured by the analysis of luminescence maps in areas without implantation. The effective depth 1 μm was estimated as the focus depth of the confocal system. In case of created NV centers, the effective depth was derived from SRIM calculations of vacancies distribution and was 0.3 μm.

Inhomogeneous spin relaxation times of single NV centers were measured using optically detected magnetic resonance technique.

## 3. Results and discussion

### 3.1. Growth-sector dependence

We found that the concentration of created NVs varied in different GSs of diamond. In inset of Fig. 2(b) one can see parts of luminescence maps of different GSs. Due to the difference in nitrogen concentration in {001} and {111} GSs, the intensity of NVs luminescence in {001} GS is lower than that in {111} GS. In addition, a step of motorized table was rather large (1 μm) and NVs in {001} GS are difficult to be registered by that way.

### 3.2. NV in {111} growth sector of HPHT sample

Due to high level of NV luminescence intensity in {111} GS we studied the smallest implanted areas in masks with a diameter of 0.5 μm (0.2 μm$^2$). It corresponds to about 20 $^4$He$^+$ ions per spot at a dose $10^{10}$ cm$^{-2}$ (100 μm$^{-2}$).

Fig. 3 shows a typical luminescence map of an implanted area and the autocorrelation function of a part of the spot. The total number of NVs in the spot is about 15. It was found that in this GS NVs tend to group in one cluster, where they cannot be optically resolved. For single photon sources creation in this sector, one would need to use lower $^4$He$^+$ ions implantation dose or smaller holes in a hardmask for implantation.

However, relatively high residual nitrogen concentration in this sector ~ $10^{17}$ cm$^{-3}$ (0.5 ppm) compared to {001} growth sector can already cause spin dephasing time ($T_2^*$) drop [29].

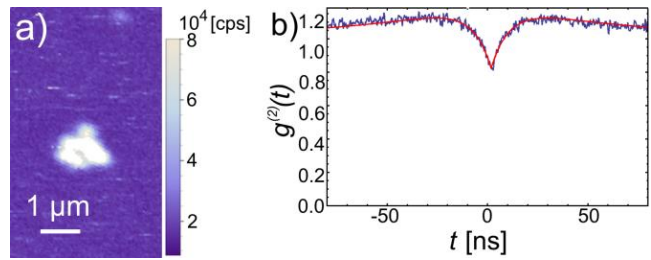

Fig. 3. a) Luminescence map of the smallest (0.5 μm diameter) implantation area with a cluster of NVs in NV03 {111} GS. The intensity is given in counts per second. b) The autocorrelation function of a part of the cluster. A number of NV centers derived from ACF is 6.

A concentration of as-grown ("intrinsic") NVs in this GS is $(2.8 \pm 0.2) \times 10^{11}$ cm$^{-3}$. This value (the order of $10^{11}$ cm$^{-3}$ or 0.5 ppt (parts-per-trillion)) is quite high and these intrinsic NVs can interfere with fabricated ones.

### 3.3. NV in {001} growth sector of HPHT sample

A concentration of as-grown NVs in this GS is $(2.8 \pm 0.1) \times 10^{10}$ cm$^{-3}$ which corresponds to an average distance between centers > 3 μm.

Fig. 4 shows a typical luminescence map of an implantation area of 3.5 μm diameter in {001} sector and the autocorrelation function of one of the spots. The average total number of NVs created in the area is 10 and most of them are single centers separated by an average distance of 1 μm. Smaller holes in masks lead to fewer NV centers (5 for 2.5 μm diameter hole and 1–2 for 1.5 μm hole) and they are still mostly single ones. We measured ACF of 1 or 2 NV centers from each implantation spot to speed up the process. The number of NV centers in the rest bright dots of an implantation spot was estimated using the luminescence intensity. For single NVs the count rate in each avalanche photodiode was $55 \pm 5$ kcounts/s. Thus, NV centers with such count rate were considered single ones. Using ACF we analyzed 31 single NV centers and the average $g^{(2)}(0)$ was $0.1 \pm 0.07$. The number of created NV centers per $^4$He$^+$ ion in our experiments was estimated to be ~ 0.01 for {001} GS and ~ 1 for {111} GS. This fact is attributed to difference in nitrogen concentration. The number of created NV centers per nitrogen atom in our experiments was estimated to be ~ $10^{-4}$ for {001} GS and ~ $10^{-3}$ for {111} GS.

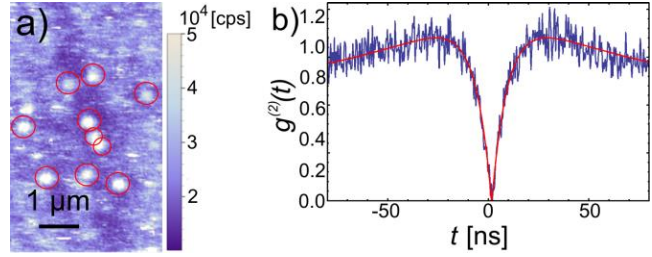

Fig. 4. a) Luminescence map of the 3.5 μm diameter implantation area with single NVs (circled in red) in NV03 {001} GS. The intensity is given in counts per second. b) The autocorrelation function of one of the single NVs.

## 3.4. ODMR measurements

To make an assessment for a NV centers coherence time we implemented an optically detected magnetic resonance (ODMR) technique measurements for single NV centers in implanted areas in NV03 diamond plate for {001} and {111} GSs. We used a permanent magnet in front of the diamond plate to separate $m_S = \pm 1$ magnetic sublevels. During the measurement, the NV centers are excited with laser, while microwave (MW) field is swept through the ODMR resonance (see Figs. 5(a) and 5(b)). Each ODMR resonance was fitted by the sum of 3 Lorentzian functions of the same width $\delta\theta$ with 2.2 MHz difference in peak positions, determined by hyperfine splitting of $m_I = 0$ and $m_I = \pm 1$ states (due to $^{14}$N nuclear spin) that results in the triplet spectral feature.

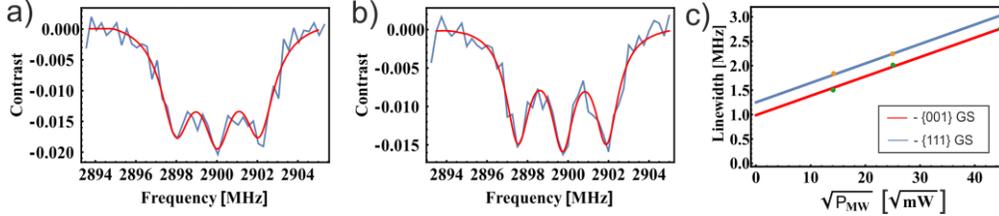

Fig. 5. a) ODMR for a single NV center in {111} GS b) ODMR for a single NV center in {001} GS c) Measured linewidth $\delta\theta$ vs. $\sqrt{P_{MW}}$ (points) and linear fit for {111} and {001} GSs.

We analyzed the $\delta\theta$ (full width half maximum) dependence on the MW power (see Fig. 5(c)) using the theoretical model $\delta\theta = \delta\theta_{MW=0} + b\sqrt{P_{MW}}$, where $\delta\theta_{MW=0}$ is the intrinsic dephasing rate, $P_{MW}$ is the applied MW power, and $b$ is a constant scaling to obtain resonance linewidth and exclude power broadening [30].

This way inhomogeneous spin relaxation time can be calculated as (see Ref. [30]):

$$T_2^* = \frac{1}{\pi \cdot \delta\theta_{MW=0}} .$$

Measured $\delta\theta_{MW=0}$ = 0.99 MHz for {001} GS and 1.25 MHz for {111} GS correspond to NV centers inhomogeneous spin relaxation times $T_2^*$ = 321 ns for {001} GS and $T_2^*$ = 254 ns for {111} GS. We attribute the difference in $T_2^*$ to the dipolar interaction between NV centers and nitrogen, which has the different concentration in GSs on the sample.

### 3.5. Discussion

We showed that HPHT IIa diamond is suitable for spatially controlled formation of single NV centers by helium ion implantation through hardmasks (vacancy-driven technique). The main idea is that $^4$He$^+$ ions locally create only vacancies which then merge with residual nitrogen atoms in substitutional position. We found that residual nitrogen concentration in {001} growth sector of such crystals ([N] ~ $10^{16}$ cm$^{-3}$) is enough for creation of several optically resolved single NV centers in the implantation spot. At the same time intrinsic [NV]/[N$_S$] (N$_S$ – substitutional nitrogen) ratio in HPHT material was found to be ~ $10^{-6}$ which leads to background NV centers concentration ~ $10^{10}$ cm$^{-3}$ (1 center in 100 μm$^3$) in low-nitrogen {001} sector. The created [NV]/[N$_S$] ratio in this sector was found to be ~ $10^{-4}$ ([NV] ~ 2 × $10^{12}$ cm$^{-3}$). This vacancy-driven technique can be preferable for some applications because it allows to independently define vacancies dose to form the desired number of NVs. Based on the average mismatch between radii of implantation holes and spots with NVs, we estimate the position accuracy as 0.5 μm, caused mostly by the migration of vacancies during annealing. Positioning accuracy could be improved by implementing fast (in-situ) annealing technique to prevent vacancies migration and annihilation without NV formation before conventional post-annealing. Since the distance between single NV centers at an implantation spot is ~ 1 μm, some of them can easily be coupled to nanowaveguides separately. Consequently, obtained results let one use this technique to create single photon sources.

It is known that after implantation nitrogen ions tend to stop in split interstitial position [31]. Such an ion can attract a vacancy and form the NV center only when two vacancies reside in second-neighbor sites. When nitrogen atom resides in a substitution position, which is preferable during the diamond growth, it needs only one vacancy near it to form the NV, in contrast to nitrogen in a split interstitial position. It means that much less vacancies (i.e. radiation damage) is needed to form the NV center when vacancy-driven technique ($^4$He$^+$ implantation) is used. Moreover, at each implantation spot we obtained several optically resolved single NVs, while after nitrogen implantation one usually gets one single NV or cluster of optically unresolved NV centers (see for example Ref. [12]).

However, single NV centers creation in CVD diamond using helium implantation with the same parameters as in our work is more challenging compared to HPHT diamond due to high intrinsic [NV]/[N$_S$] ~ $10^{-3}$ ratio [32]. Thus, for a sample with nitrogen content ~ 10 ppm, intrinsic NV density is ~ 10 ppb [33]. Therefore, to obtain single NV centers in predetermined positions one needs significantly lower intrinsic nitrogen concentration [34].

The obtained inhomogeneous spin relaxation times ~ 300 ns are close to values measured using ODMR linewidth method for HPHT ($^4$He$^+$ implantation) and CVD diamonds (electron irradiation) with nitrogen content < 1 ppm [10], [16]. In isotopically purified $^{12}$C CVD diamond with close nitrogen content (~ 1 ppm), however, ODMR signal can be five times narrower (~ 200 kHz) both for $^4$He$^+$ implantation and electron irradiation [35], [33]. This suggests that inhomogeneous spin relaxation times obtained in this work can be further improved utilizing isotopically purified $^{12}$C HPHT diamond.

### 4. Conclusion

We created NV centers by ion implantation through hardmasks in {111} and {001} growth sectors of IIa diamond grown by temperature gradient technique at high-pressure high-temperature (HPHT). It was found, that in {111} GS of HPHT diamond clusters of NVs are usually created, while in {001} GS of HPHT sample single NV centers are produced. NV centers inhomogeneous spin relaxation time was assessed to be in the order of 300 ns.

Thus, we consider the purest {001} GS of IIa-type colorless HPHT crystal to be a good material for single NV centers formation for applications in quantum photonics. Due to its quite low nitrogen concentration, extremely low concentration of intrinsic NV centers and high crystalline quality, it is suitable both for common $N^+$ ion implantation process and for vacancy-driven process of NV creation based on $He^+$, $C^+$ implantation or $e^–$ and fs-laser pulses irradiation.

**Appendix A: Nitrogen concentration measurements**

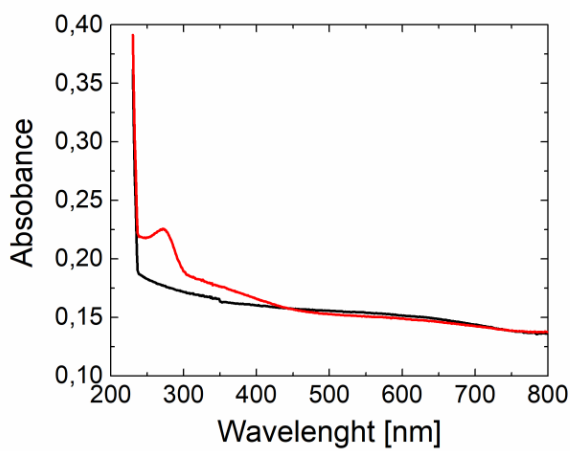

Fig. 6. Ultraviolent absorption of NV03 in {001} growth sector (red) and in {001} growth sector (black).

In Fig. 6, ultraviolet absorption data of different growth sectors of NV03 are shown. Using the graph and accounting for the sample thickness of 0.55 mm, one can derive the nitrogen content in the diamond crystal using the formula $[N] = 0.56\alpha$, where $\alpha$ (cm$^{-1}$) is absorption coefficient at 270 nm and $[N]$ (ppm) is the nitrogen concentration [19]. The nitrogen concentration in {111} growth sector was ~ 500 ppb. Since in {001} growth sector the nitrogen content was undetectable via this method, we estimated the nitrogen concentration in {001} growth sector < 50 ppb (at least one order lower than in {111}).

## Funding

The luminescence mapping and NV centers properties measurements were supported by Russian Science Foundation (RSF) (18-72-00232); Ministry of Education and Science of the Russian Federation (FNNR-2019-0004).

## Acknowledgements

We thank Taisia E. Drozdova for nitrogen concentration measurements.

## Disclosures

The authors declare no conflicts of interest.